\documentclass{ws-procs975x65}

\newcommand{\be}{\begin{equation}}
\newcommand{\ee}{\end{equation}}
\newcommand{\bel}[1]{\begin{equation}\label{#1}}
\newcommand{\ba}{\begin{eqnarray}}
\newcommand{\ea}{\end{eqnarray}}
\newcommand{\bal}[1]{\begin{eqnarray}\label{#1}}

\begin{document}

\title{DETECTING COALESCENCES OF INTERMEDIATE-MASS BLACK HOLES IN GLOBULAR CLUSTERS\\ WITH THE EINSTEIN TELESCOPE}

\author{I. MANDEL}
\address{NSF Astronomy and Astrophysics Postdoctoral Fellow\\
Department of Physics and Astronomy, Northwestern University\\
Evanston, IL 60208\\
\email{ilyamandel@chgk.info}
}

\author{J. R. GAIR}
\address{Institute of Astronomy, University of Cambridge\\
Cambridge, CB30HA, UK}

\author{M. C. MILLER}
\address{Department of Astronomy and Center for Theory and Computation, University of Maryland\\
College Park, MD 20742}

\begin{abstract}

We discuss the capability of a third-generation ground-based detector such as the Einstein Telescope (ET) to detect mergers of intermediate-mass black holes (IMBHs) that may have formed through runaway stellar collisions in globular clusters.  We find that detection rates of $\sim 500$ events per year are plausible \cite{Gair:2009ETrev}.

\end{abstract}

\keywords{Gravitational Waves; Intermediate-Mass Black Holes; the Einstein Telescope.\\ \\}

\bodymatter

The Einstein Telescope (ET), a proposed third-generation ground-based gra\-vi\-ta\-tio\-nal-wave (GW) detector, will be able to probe GWs in a frequency range reaching down to $\sim 1$ Hz 
\cite{Freise:2009}. This bandwidth will allow the ET to probe sources with masses of hundreds or a few thousand $M_\odot$ which are out of reach of LISA 
or the current ground-based detectors LIGO, Virgo, and GEO-600. 

Globular clusters may host intermediate-mass black holes (IMBHs) with masses in the $\sim 100$ -- $1000\ M_\odot$ range (see Ref.~\refcite{Miller:2009} and references therein). 
If the stellar binary fraction in a globular cluster is sufficiently high, two or more IMBHs can form \cite{Fregeau:2006}.  These IMBHs then sink to the center in a few million years, where they form a binary and merge via three-body interactions with cluster stars followed by gravitational radiation reaction (see \cite{Fregeau:2006,Amaro:2007} for more details).
Therefore, the rate of IMBH binary mergers is just the rate at which pairs of IMBHs form in clusters. The rate of detectable coalescences is
\bel{IMBHIMBHrategeneral}
R \equiv \frac{dN_{\rm event}}{dt_o}=
\int_{M_{\rm tot, min}}^{M_{\rm tot, max}} dM_{\rm tot}
\int_0^1 dq \int_0^{z_{\rm max}(M_{\rm tot},q)} dz 
\frac{d^4 N_{\rm event}} {dM_{\rm tot} dq dt_e dV_c} \frac{dt_e}{dt_o}  \frac{dV_c}{dz}.
\ee
Here 
$M_{\rm tot}$ is the total mass of the coalescing IMBH-IMBH binary and $q\leq 1$ is the mass ratio between the IMBHs;
$z_{\rm max}(M_{\rm tot},q)$ is the maximum redshift to which
the ET could detect a merger between two IMBHs of total mass $M_{\rm tot}$ and mass ratio $q$;
$dt_e/dt_o=(1+z)^{-1}$ is the relation between local time and our observed time,
and $dV_c/dz$ is the change of comoving volume with redshift, given by
\bel{Vc}
\frac{dV_c}{dz}=4\pi D_H^3 \left[\Omega_M(1+z)^3+\Omega_\Lambda\right]^{-1/2}
\left\{\int_0^z\frac{dz^\prime}{\left[\Omega_M(1+z^{\prime})^3+
\Omega_\Lambda\right]^{1/2}}\right\}^2.
\ee
We assume a flat universe ($\Omega_k=0$), and use $\Omega_M=0.27$,
$\Omega_\Lambda=0.73$, $H_0=72$~km~s$^{-1}$~Mpc$^{-1}$, and $D_H=c/H_0\approx 4170$~Mpc, so that the luminosity distance can be written as a function of redshift as \cite{Hogg:1999}:
\bel{DLz}
D_L(z) = D_H (1+z) \left\{\int_0^z\frac{dz^\prime}{\left[\Omega_M(1+z^{\prime})^3+
\Omega_\Lambda\right]^{1/2}}\right\}.
\ee

We make the following assumptions.  {\bf 1.} IMBH pairs form in a fraction $g$ of all globular clusters.
{\bf 2.} We neglect the delay between cluster formation and IMBH coalescence.
{\bf 3.} When an IMBH pair forms in a cluster, its total mass is a fixed fraction of the cluster mass,
$M_{\rm tot}=2\times 10^{-3}~M_{\rm cl}$, consistent with simulations \cite{Gurkan:2004}.  The mass ratio is uniform in $[0,1]$.  We restrict our attention to systems with a total mass between $M_{\rm tot, min}=100 M_\odot$ and $M_{\rm tot, max}=20000 M_\odot$.
Thus,
\be 
\frac{d^4 N_{\rm event}} {dM_{\rm tot} dq dt_e dV_c} = g \frac{d^3 N_{\rm cl}} {dM_{\rm cl} dt_e dV_c} \frac{1}{2\times 10^{-3}}.
\ee
{\bf 4.} The distribution of cluster masses scales as $(dN_{\rm cl}/dM_{\rm cl}) \propto M_{\rm cl}^{-2}$ independently of redshift.  We confine our attention to clusters with masses ranging from  $M_{\rm cl, min}=5\times10^4 M_\odot$ to $M_{\rm cl, max}=10^7 M_\odot$.  The total mass formed in all clusters in this mass range at a given redshift 
is a redshift-independent fraction $g_{\rm cl}$ of the total star formation rate per comoving volume:
\begin{equation}
\frac{d^3 N_{\rm cl}} {dM_{\rm cl} dt_e dV_c}= \frac{g_{\rm cl}}{\ln (M_{\rm cl, max}/M_{\rm cl, min})} \frac{d^2M_{\rm SF}}{dV_c dt_e} \frac{1}{M_{\rm cl}^2}.
\end{equation}
{\bf 5.} The star formation rate as a function of redshift $z$  rises rapidly with increasing $z$ to $z\sim 2$,
after which it remains roughly constant \cite{Steidel:1999}:
\begin{equation}
\frac{d^2M_{\rm SF}}{dV_c dt_e}=0.17\frac{e^{3.4z}}{e^{3.4z}+22}
\frac{\left[\Omega_M(1+z)^3+\Omega_\Lambda\right]^{1/2}}
{(1+z)^{3/2}}~M_\odot~{\rm yr}^{-1}~{\rm Mpc}^{-3}.
\end{equation}


Rather than computing $z_{\rm max}(M_{\rm tot},q)$ [Eq.~\ref{IMBHIMBHrategeneral}] for all values of $M_{\rm  tot}$ and $q$, we rely on the following fitting formula for the luminosity-distance range $D_{\rm L, max}$ as a function of the redshifted total mass $M_z=M_{\rm tot} (1+z)$, obtained by using the effective-one-body, numerical relativity (EOBNR) gravitational waveforms \cite{Buonanno:2007EOBNR} to model the inspiral, merger, and ringdown phases of coalescence:
\be
D (M_z) = (A \ {\rm Mpc})   
	\left\{ \begin{array}{ll}
	(M_z/M_\odot)^{3/5}&\mbox{if } M_z<M_0 \\
	(M_0/M_\odot)^{11/10} (M_z/M_\odot)^{-1/2}&\mbox{if }  M_z>M_0
	\end{array} \right. ,
\ee
where $A=500$, $M_0=600 M_\odot$ for $q=1$ and $A=281$, $M_0=450 M_\odot$ for $q=0.25$.  We use $\rho=8$ as the SNR threshold for a ``single ET'' configuration.  We determine the sky-location and orientation averaged range by dividing the horizon distance by $2.26$\cite{FinnChernoff:1993}, ignoring redshift corrections to this factor.

We can compute $z(D_L)$ by inverting Eq.~(\ref{DLz}).  For a given choice of $M_{\rm tot}$ and $q$, the maximum detectable redshift $z_{\rm max} (M_{\rm tot}, q)$ is then obtained by finding a self-consistent solution of 
$z\Big(D_{\rm L, max}\big(M_{\rm tot} (1+z_{\rm max})\big)\Big)=z_{\rm max}.$

In order to compute the rate of detectable coalescences, we carry out the integrals over $M_{\rm tot}$ and $z$ in Eq.~(\ref{IMBHIMBHrategeneral}) for two specific values of $q$.  For $q=1$, we find the total rate to be $R=7.5\times10^4\ g\ g_{\rm cl}$ yr$^{-1}$; for $q=0.25$, it is $R=2.7\times10^4\ g\ g_{\rm cl}$ yr$^{-1}$.  The range varies smoothly with $q$; therefore, we estimate that full rate, including the integral over $q$ is
\bal{IMBHIMBHrate}
R &=& \frac{2\times 10^{-3} \  g \  g_{\rm cl}\ {\rm yr}^{-1}} {\ln (M_{\rm tot, max}/M_{\rm tot, min})} 
\int_{M_{\rm tot, min}}^{M_{\rm tot, max}}  \frac{M_\odot dM_{\rm tot}}{M_{\rm tot}^2} \int_0^1 dq\\
\nonumber 
&&
\int_0^{z_{\rm max}(M_{\rm tot},q)} dz \ 
0.17\frac{e^{3.4z}}{e^{3.4z}+22}
\frac{4\pi (D_H/{\rm Mpc})^3} {(1+z)^{5/2}}
\times
\left\{\int_0^z\frac{dz^\prime}{\left[\Omega_M(1+z^{\prime})^3+
\Omega_\Lambda\right]^{1/2}}\right\}^2\\
\nonumber
&\approx& 500 \left(\frac{g}{0.1}\right) \left(\frac{g_{\rm cl}}{0.1}\right) {\rm yr}^{-1},
\ea
where we arbitrarily chose $g=0.1$ and $g_{\rm cl}=0.1$ as the default scalings.

Mergers between pairs of globular clusters containing IMBHs can increase this rate by up to a factor of $\sim 2$ \cite{AmaroSeoaneSantamaria:2009}.  Ref.~\refcite{Gair:2009ETrev} contains additional details on coalescences involving intermediate-mass black holes as gravitational-wave sources for the ET.

\section*{Acknowledgments}

IM is supported by the NSF Astronomy and Astrophysics Postdoctoral Fellowship under award AST-0901985 and was partially supported from NASA ATP Grant NNX07AH22G.  JG's work is supported by a Royal Society University Research Fellowship.  MCM acknowledges NASA ATP grant NNX08AH29G.  IM's participation in MG12 was enabled by an NSF travel grant.

\bibliographystyle{ws-procs975x65}
\bibliography{main}

\begin{thebibliography}{10}

\bibitem{Gair:2009ETrev}
J.~R. {Gair}, I.~{Mandel}, M.~C. {Miller} and M.~{Volonteri}, {\em ArXiv
  e-prints}  (2009), 0907.5450.

\bibitem{Freise:2009}
A.~{Freise}, S.~{Chelkowski}, S.~{Hild}, W.~{Del Pozzo}, A.~{Perreca} and
  A.~{Vecchio}, {\em Classical and Quantum Gravity} {\bf 26}, 085012 (2009).

\bibitem{Miller:2009}
M.~C. {Miller}, {\em ArXiv e-prints}  (2008), 0812.3028.

\bibitem{Fregeau:2006}
J.~M. {Fregeau}, S.~L. {Larson}, M.~C. {Miller}, R.~{O'Shaughnessy} and F.~A.
  {Rasio}, {\em Astrophysical Journal Letters} {\bf 646}, L135 (2006).

\bibitem{Amaro:2007}
P.~{Amaro-Seoane}, J.~R. {Gair}, M.~{Freitag}, M.~C. {Miller}, I.~{Mandel},
  C.~J. {Cutler} and S.~{Babak}, {\em Classical and Quantum Gravity} {\bf 24},
  113 (2007).

\bibitem{Hogg:1999}
D.~W. {Hogg}, {\em ArXiv Astrophysics e-prints}  (1999), astro-ph/9905116.

\bibitem{Gurkan:2004}
M.~A. {G{\"u}rkan}, M.~{Freitag} and F.~A. {Rasio}, {\em \apj} {\bf 604}, 632
  (2004).

\bibitem{Steidel:1999}
C.~C. {Steidel}, K.~L. {Adelberger}, M.~{Giavalisco}, M.~{Dickinson} and
  M.~{Pettini}, {\em \apj} {\bf 519}, 1 (1999).

\bibitem{Buonanno:2007EOBNR}
A.~{Buonanno}, Y.~{Pan}, J.~G. {Baker}, J.~{Centrella}, B.~J. {Kelly}, S.~T.
  {McWilliams} and J.~R. {van Meter}, {\em \prd} {\bf 76}, 104049 (2007).

\bibitem{FinnChernoff:1993}
L.~S. {Finn} and D.~F. {Chernoff}, {\em \prd} {\bf 47}, 2198 (1993).

\bibitem{AmaroSeoaneSantamaria:2009}
P.~{Amaro-Seoane} and L.~{Santamaria}, {\em ArXiv e-prints}  (2009), 0910.0254.

\end{thebibliography}

\end{document}